# GreenNIDS: A Benchmark Workflow and Analysis Dataset for NetFlow Intrusion Detection Experiments

Lilia-Eliana Popescu-Apreutesei* and Sabina-Cristiana Necula*

Faculty of Economics and Business Administration,
Alexandru Ioan Cuza University of Iasi, 700505 Iasi, Romania
* Corresponding authors
lilia.popescu@student.uaic.ro
sabina.necula@uaic.ro

Abstract. Comparing the predictive performance and computational cost of flow-based intrusion detection experiments requires both quantities to be available for each run. GreenNIDS connects benchmark execution on public NetFlow datasets to a reusable analysis dataset. Model implementations write a common set of outputs, which are ingested into a run-level table with companion tables for model configuration, feature importance, and training history. The run-level table combines predictive metrics with timing, throughput, memory use, and energy estimates based on thermal design power. A CPU-only demonstration comprises 216 runs across four dataset families, three versions, and binary and multiclass tasks. Ingestion produces a 216 × 96 analysis table with no duplicate run identifiers and complete Matthews correlation coefficient (MCC) and F1 coverage. Multiclass ROC-AUC is missing for 36 PyTorch runs, making metric availability an explicit part of subsequent comparisons. The stored records support grouped performance and training-time plots, as well as resource screening after execution. Combining task-specific MCC thresholds with limits on training time, peak RAM, and estimated training energy retains 25.0% of the full run collection. These results establish compatibility within the evaluated collection and support comparisons after execution, subject to the recorded preprocessing, execution profile, and approximate energy accounting.

Keywords: Network intrusion detection, NetFlow, benchmarking, Green AI, reproducibility

## 1 Introduction

Predictive scores alone cannot establish whether a network intrusion detection experiment can be repeated within a given computational budget. Training time determines how many experiments fit into an allocation, while peak memory use can determine whether a run fits at all. The Green AI position paper argues for evaluating computational cost alongside predictive performance [1], and Henderson et al. provide methods for retaining energy and carbon information with machine learning experiments [2].

For a NetFlow benchmark, this motivates keeping resource use and predictive results together at the level of each run.

Such comparisons also depend on what a dataset contains and how it is processed. Reviews of intrusion detection datasets document differences in traffic collection, attack coverage, and data representation [3, 4]. Sarhan et al. provide NetFlow representations of several source corpora [5], while their subsequent comparison of feature sets examines performance across different network environments and attack scenarios [6]. A shared flow representation supplies a common basis for experimentation, but each result still belongs to a particular dataset version and preprocessing procedure. Retaining these identifiers is necessary to interpret differences between runs.

Reusing experiments requires more than retaining their final scores. OpenML associates results with tasks, algorithm implementations, and parameter settings [7], while MLflow records parameters and metrics together with experiment artifacts [8]. These systems establish approaches to experiment tracking that can inform a NetFlow benchmark. The application developed here connects dataset-specific preparation to a common set of run outputs and a tabular representation for subsequent analysis. Kouper and Stone document limited public availability of cybersecurity research data [9]. Fisk et al. describe privacy and organisational disclosure risks in sharing network traffic and logs [10]. Public benchmark corpora provide an accessible setting in which to document the connection between data preparation and reusable experiment records.

GreenNIDS links the outputs of tree-based estimators and neural networks through a common analysis schema. The study examines whether that schema accommodates the evaluated implementations and which comparisons their records support after execution. Two research questions organise the study.

**RQ1:** Can heterogeneous outputs from the evaluated NetFlow model implementations be represented in a common run-level schema with linked companion tables?

**RQ2:** Which kinds of comparative analysis become available once predictive metrics, timing, memory, throughput, energy estimates, and post-ingestion screening fields are stored in one benchmark table?

## 2 Related Work

Intrusion detection evaluation depends on the relationship between experimental traffic and the environment in which a detector would operate. Sommer and Paxson examine how traffic variability, the cost of errors, and the interpretation of anomalies complicate machine learning for network intrusion detection [11]. Evaluation design introduces further choices. Milenkoski et al. organise intrusion detection evaluation around workloads and measurement methods [12], while Arp et al. identify methodological pitfalls throughout security machine learning, including biased data and inappropriate evaluation [13]. These studies motivate retaining the experimental conditions needed to interpret a reported score.

Dataset standardisation addresses part of this problem. The NetFlow datasets introduced by Sarhan et al. give several source corpora a common representation [5]. Their later comparison of NetFlow and CICFlowMeter features examines how the choice of representation affects classification across datasets [6]. Dataset surveys address a broader set of concerns, including the relationship between traffic generation and the intended evaluation task [3, 4]. GreenNIDS retains dataset family and version as grouping variables so that runs from different releases remain distinguishable when results are aggregated.

System comparisons and the reuse of individual experiments place different demands on benchmarking. MLPerf Training and MLPerf Inference specify workloads and evaluation procedures for comparing systems under defined conditions [14, 15]. OpenML and MLflow preserve individual experiments by linking their outcomes to configurations and saved outputs [7, 8]. A later analysis needs access to those records as well as an account of how the comparison was conducted. Gaps in experimental documentation identified by Gundersen and Kjensmo [17] explain why retaining that account matters. The reporting checklists and replication activities examined by Pineau et al. [18] provide ways to make published experiments open to checking.

Energy comparisons also depend on how the reported quantity was obtained. MLPerf Power defines measurement procedures across hardware platforms [16], while Green Algorithms estimates computational emissions from hardware and runtime information [19]. Henderson et al. describe systematic reporting of energy and carbon footprints [2], and Bouza et al. compare estimation tools and their underlying assumptions [20]. These distinctions determine how energy values can be interpreted. GreenNIDS retains its energy fields as TDP-based estimates, with their measurement limitations considered in the Discussion.

GreenNIDS brings these requirements to a NetFlow workload by connecting dataset preparation to run records that can be queried after execution. The evaluation examines whether outputs from the included implementations share one representation and whether the retained measurements support grouped comparisons and resource screening.

## 3 GreenNIDS Benchmark

GreenNIDS separates model execution from the analysis of completed experiments (Fig. 1). During execution, dataset preparation and model training produce outputs in a common format. Ingestion then reads those outputs into related tables (Table 1), allowing analyses to use named fields without interpreting the original folder structure for every comparison. Model-specific training code and NetFlow-specific preparation remain separate from the definitions used to organise results.

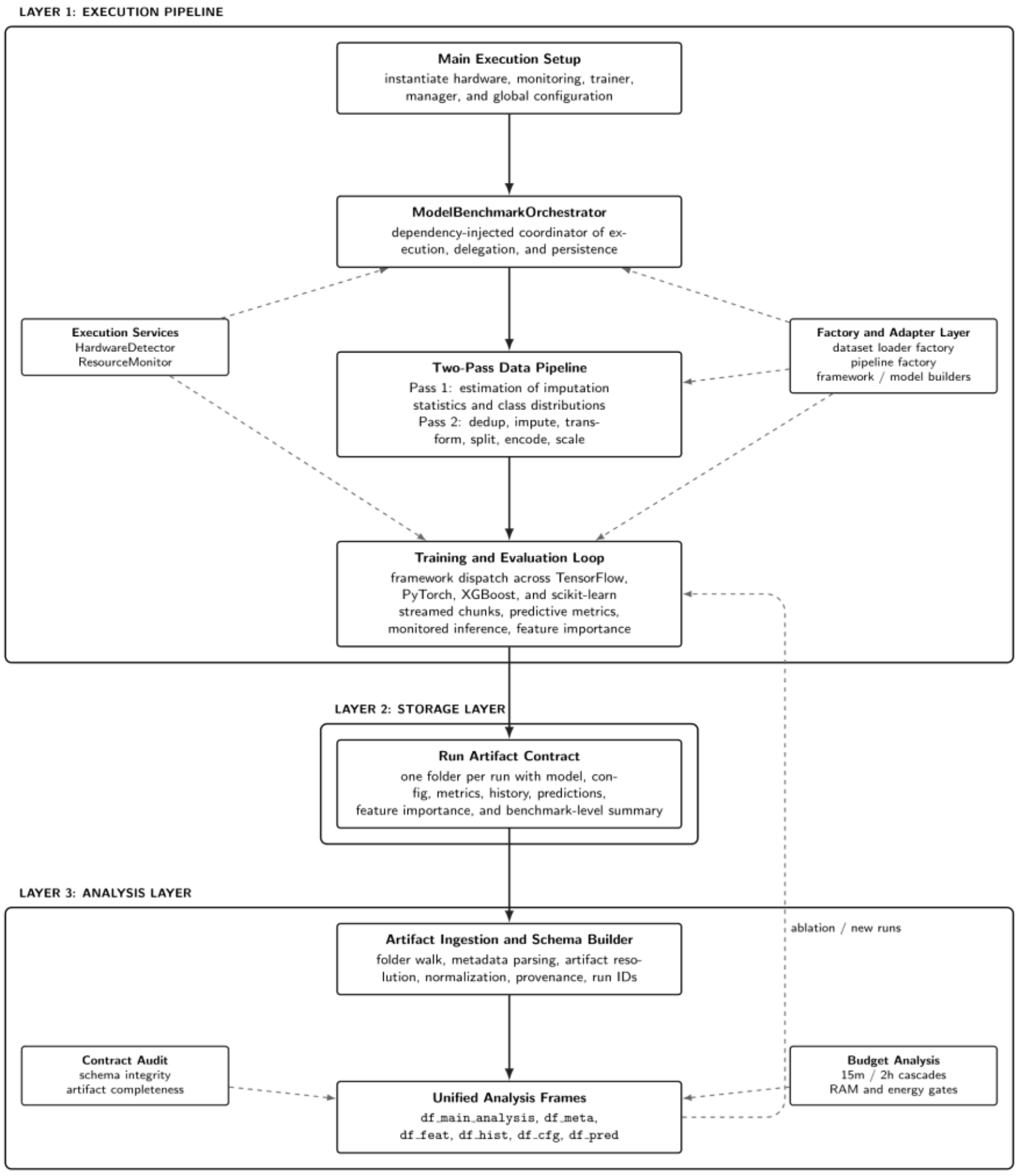

**Fig. 1.** GreenNIDS execution, storage, and analysis components. Detailed model outputs are linked to the analysis tables through a shared run identifier.

Separating model execution from resource monitoring allows the same measurements to accompany implementations from different libraries. ModelBenchmarkOrchestrator coordinates data preparation and output writing, with ModelTrainer handling training and evaluation. HardwareDetector records the execution environment and device thermal design power (TDP), while ResourceMonitor samples elapsed time, RAM use, and CPU load. GPU memory and utilisation are available on supported execution paths. ModelManager saves the resulting files and delegates plotting to ModelVisualizer. These outputs supply the common input to ingestion.

Models enter through a registry that associates each estimator with its construction procedure and configuration. HistGradientBoosting, LightGBM, and XGBoost use parameter dictionaries. TensorFlow networks are constructed through create_tf_nn, and PyTorch networks through get_pytorch_models. The orchestrator applies hardware-dependent settings where required and sends each model through the common evaluation and output-writing procedures. Adding an estimator requires a compatible registration and adapter so that its results follow the existing output format.

Dataset preparation uses a registry indexed by family and version. Naming variants for the NF-CICIDS2018 family are resolved before aggregation. Large CSV files are processed in chunks. The first pass computes or loads median imputation values from a source-data sample. The second handles labels and feature exclusions, replaces non-finite values, removes duplicates within chunks, applies imputation, and selectively log-transforms quantity features. A stratified sample is then assembled and split into training and test sets. After splitting, classes with fewer than two training examples or one test example are excluded, and multiclass labels are re-encoded to contiguous integers. RobustScaler is fitted on the training data and applied to the test data. Imputation therefore precedes the final split, while scaling uses training data only.

**Table 1.** GreenNIDS dataframe family and grouped structure of df_main_analysis.

| Frame | Shape | Key | Role and stored content |
|---|---|---|---|
| *df_main* | $N \times p_{\text{main}}$ | Run_ID | Master record containing one row per run, summary metrics, JSON-derived fields, and artifact paths. |
| *df_main_analysis* | $N \times p_{\text{analysis}}$ | Run_ID | Run-level measurements and execution metadata used for grouped comparisons and resource screening. Column groups are listed below. |
| *df_meta* | $N \times p_{\text{meta}}$ | Run_ID | Artifact paths and raw JSON payloads linking the recorded measurements to their saved files. |
| *df_feat* | $\left(\sum_{r=1}^{N} F_r\right) \times p_{\text{feat}}$ | Run_ID | Feature names and importance values linked to run identifiers and summary metrics. |
| *df_hist* | $\left(\sum_{r=1}^{N} H_r\right) \times p_{\text{hist}}$ | Run_ID | Training-history steps and their associated metrics. |
| *df_cfg* | $N \times p_{\text{cfg}}$ | Run_ID | Model configuration, architecture tags, and extracted parameter values. |
| *df_pred* | $\left(\sum_{r=1}^{N} P_r\right) \times p_{\text{pred}}$ | Run_ID | Optional prediction records, available when prediction export is enabled and its files are loaded. |

Notation: N is the number of runs and p denotes the number of columns in the indicated frame. $F_r$, $H_r$, and $P_r$ are the feature-importance, history, and prediction rows contributed by run r. The 96 columns in df_main_analysis comprise identity (18), architecture (2), quality semantics (6), quality metrics (19), system performance (9), resource use and Green AI proxy (8), training behaviour (9), feasibility and completeness (15), and reproducibility (10). Reported dimensions and metric availability are given in Table 2.

Each run produces a model binary and JSON files containing its configuration and metrics. Training history and feature importance are stored in Parquet when available, and prediction export is optional. Ingestion reads the run folders and constructs the dataframe family in Table 1. Run_ID is a hash of the experiment, version, task, dataset, model, framework, accelerator, and base filename. The same key links summary metrics to the companion tables. Its reproducibility therefore depends on preserving these identifying values, including the base filename. Paths to the artifacts are retained separately in df_meta, and df_main preserves the full master record.

The analysis schema is versioned as 3.0, and the main table contains one row per run. Companion tables accommodate outputs with different row counts, such as feature-importance entries and training-history steps, without duplicating the run-level record. Model-configuration fields remain available for selecting comparable subsets. This separation lets an analysis use summary metrics alone or join them to the detailed outputs required by a particular question.

The analysis table preserves run-level quantities alongside derived screening indicators. Resource limits can consequently be changed by recomputing indicators from the retained measurements. A new analytical variable can be obtained without retraining only when the stored outputs contain the information it requires. NetFlow-specific registration, label handling, and preprocessing would need adaptation before this workflow could be used for other tabular-classification tasks.

## 4 Experimental Setup and Analysis Dataset

All 216 runs used a CPU-only notebook environment. The hardware record reports Linux x86_64, 2 physical CPU cores, 4 logical CPU cores, and 31.35 GB RAM, with no detected GPU or TPU. The energy estimator used a CPU TDP value of 65 W. Recorded software versions were TensorFlow 2.18.0, PyTorch 2.6.0+cu124, XGBoost 2.0.3, scikit-learn 1.2.2, pandas 2.2.3, NumPy 1.26.4, Matplotlib 3.7.2, and Seaborn 0.12.2. The global seed was 42. Deterministic mode was disabled and mixed precision was not used. GPU execution is supported in the benchmark code, but the present comparisons concern the recorded CPU environment.

The run collection covers the BoT-IoT, CSE-CIC-IDS2018, ToN-IoT, and UNSW-NB15 NetFlow families. Three versions of each family give twelve dataset instances, each evaluated for binary and multiclass classification. The nine model configurations comprise HistGradientBoosting, LightGBM, XGBoost, three TensorFlow multilayer perceptrons, and three PyTorch multilayer perceptrons. Each version–task combination contains 12 tree-based runs and 24 neural-network runs. The neural networks carry Small, Medium, and Large identifiers within each implementation, giving 12 runs per identifier and task. These identifiers provide grouping labels for the recorded configurations.

The experiments use a stratified 80/20 train–test split with seed 42 and a shared preprocessing configuration. No hyperparameter optimisation was performed. Recorded input dimensions are 10 for v1, 41 for v2, and 51 for v3, referring to the features supplied to the models after preparation. Screening indicators use MCC thresholds of 0.70 for binary classification and 0.60 for multiclass classification, training-time limits of 15 minutes and 2 hours, an 8 GB peak-RAM limit, and a 5 Wh estimated training-energy limit. These are analyst-selected thresholds for examining the run collection. They are computed after execution and can be recalculated without altering the trained models.

The unit of analysis is an executed dataset–task–model configuration. Balanced coverage makes each version–task group comparable in its number of tree and neural runs, but runs on different datasets answer different prediction problems. Grouped summaries are used descriptively. The reported design supplies point estimates for the selected split and execution environment.

## 5 Results

For RQ1, all 216 runs fit one analysis schema with unique identifiers and linked companion tables. MCC and F1 are available for every run. Both artifact-completeness indicators reach 100.0%, although expected AUC values are missing for some runs. Table 2 reports dimensions, metric availability, and screening rates.

**Table 2.** Reported ingestion results, AUC missingness, and screening rates across all dataset versions. MCC pass rates use the 108 runs for the corresponding task. Other percentages use all 216 runs.

| Audit item | Value |
|---|---|
| df_main shape | 216 × 107 |
| df_main_analysis shape | 216 × 96 |
| df_meta shape | 216 × 12 |
| df_feat shape | 6,480 × 23 |
| df_hist shape | 9,269 × 29 |
| df_cfg shape | 216 × 24 |
| Duplicate Run_ID rows | 0 |
| ROC_AUC_Filled null rate | 36 rows (16.7%) |
| ROC_AUC_Binary null rate | 108 rows (50.0%) |
| ROC_AUC_OVR_Macro null rate | 144 rows (66.7%) |
| ROC_AUC_OVR_Weighted null rate | 144 rows (66.7%) |
| MCC_Pass_Binary | 93.5% |
| MCC_Pass_Multi | 86.1% |
| Time_Budget_15m_Pass | 61.6% |
| Time_Budget_2h_Pass | 100.0% |
| RAM_Budget_Pass | 63.0% |
| Energy_Budget_Pass | 35.6% |
| Core_Budgets_Pass | 37.5% |
| All_Budgets_Pass | 25.0% |
| Run_Is_Complete_Core | 100.0% |
| Run_Is_Complete_Full | 100.0% |

Missing AUC values have different meanings across tasks. The binary AUC column has 108 null entries, corresponding to multiclass runs for which that field is inapplicable. Each multiclass one-versus-rest column has 144 null entries: the 108 binary runs and 36 PyTorch multiclass runs with missing AUC. The combined ROC_AUC_Filled field has 36 null entries, all associated with those PyTorch multiclass runs. The PyTorch inference path omits the probability conversion required by the AUC interface. Those runs remain available for MCC and F1 analyses but are excluded from AUC summaries, so an AUC comparison uses a different subset from an MCC comparison.

Figure 2 compares predictive performance and training time within dataset families and versions, separately for binary and multiclass classification. Individual points retain the variation among model configurations, while diamonds show medians for the tree-based and neural-network groups. This grouping keeps the dataset context visible when comparing the computational cost associated with predictive scores. MCC is used because its binary formulation incorporates all four confusion-matrix outcomes [21], and its extension permits evaluation with more than two classes [22]. The analysis table also retains accuracy, precision, recall, F1, and available AUC values for alternative summaries.

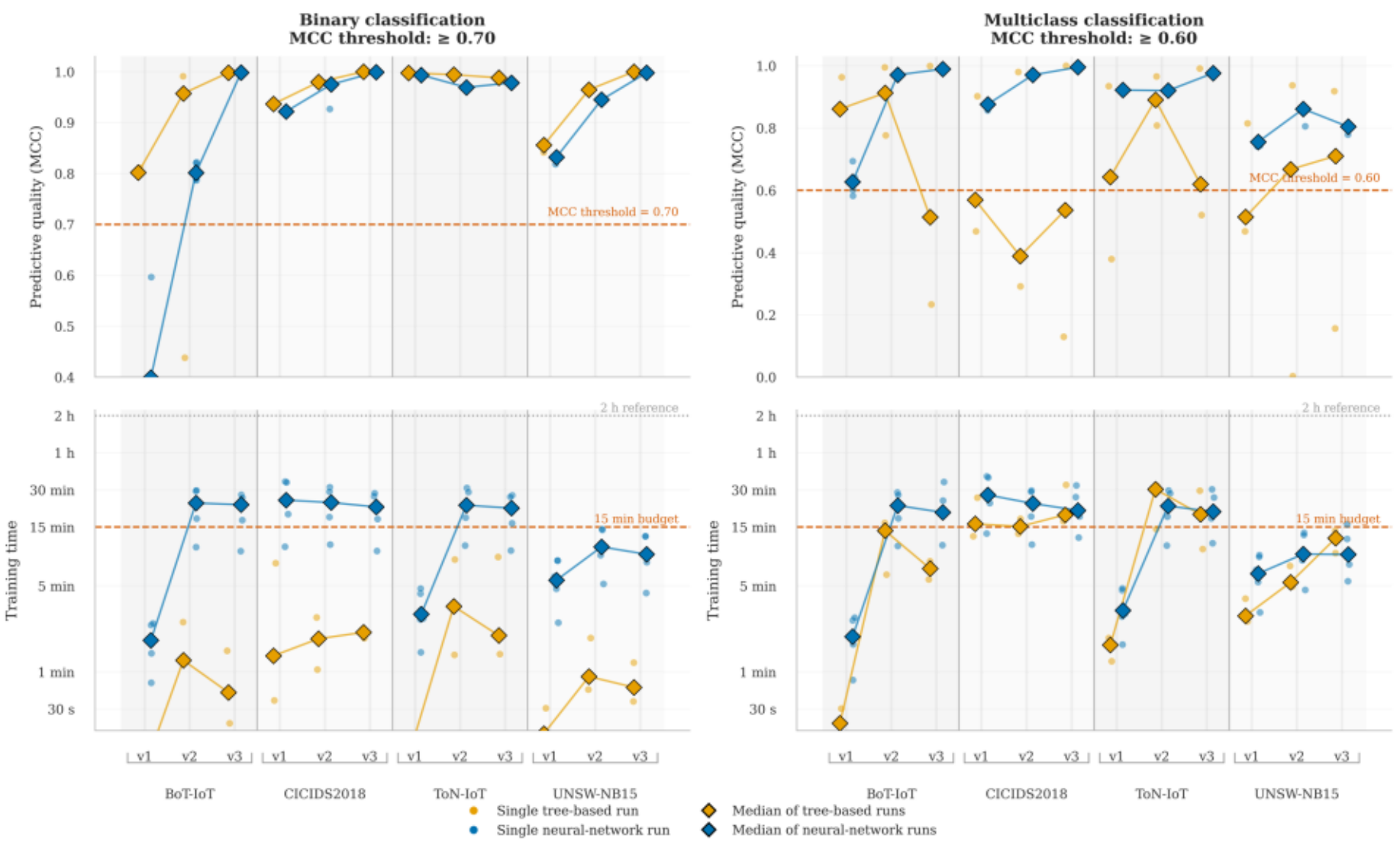

**Fig. 2.** MCC and training time by dataset family, version, and classification task. Faint points represent individual runs. Diamonds show medians for tree-based and neural-network groups. Dashed lines mark the selected MCC and 15-minute thresholds, with a 2-hour reference in the timing panels.

The individual resource criteria retain different proportions of the full 216-run collection. Every run meets the 2-hour training limit, while 61.6% meet the 15-minute limit. The 8 GB RAM criterion retains 63.0%, and the 5 Wh estimated training-energy criterion retains 35.6%. Applying the task-specific MCC threshold together with the 15-minute and 8 GB limits retains 37.5%. Adding the energy criterion leaves 25.0%. These joint proportions describe the intersection of the specified conditions across all three dataset versions.

Figure 3 applies the cumulative criteria within v2 only, with 12 tree-based runs or 24 neural-network runs as the denominator for each task. The curves locate the stage at which runs leave each group. Their percentages describe those v2 groups, whereas Table 2 summarises all three dataset versions.

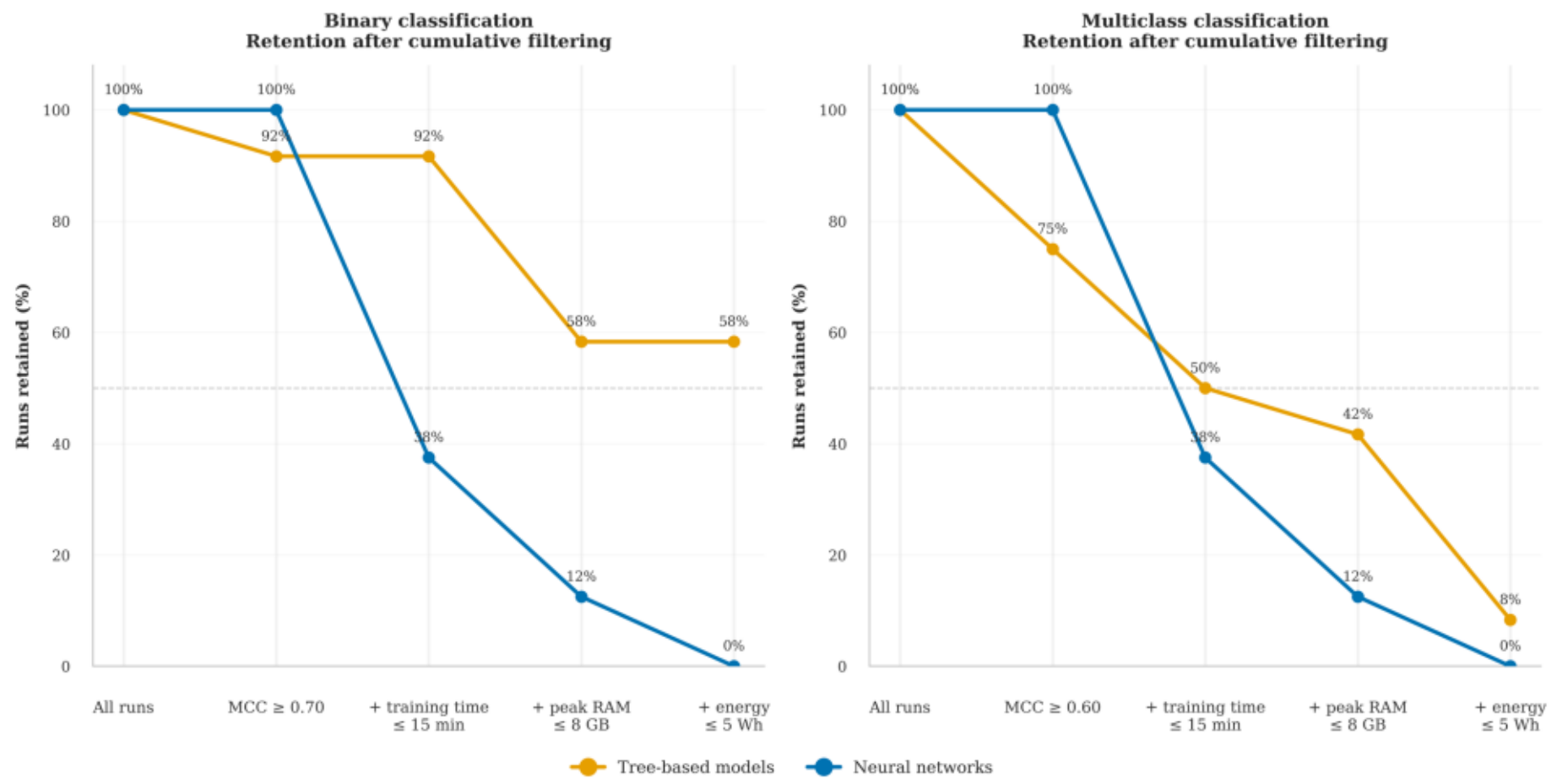

**Fig. 3.** Cumulative screening of v2 runs only, separated by task and model family. Each curve begins with its own group at 100%. Successive stages apply the task-specific MCC threshold, a 15-minute training limit, an 8 GB peak-RAM limit, and a 5 Wh estimated training-energy limit. Percentages are rounded to whole numbers.

## 6 Discussion and Limitations

RQ1 is answered within the evaluated collection: the outputs of its tree-based and neural-network implementations fit one run-level schema, and Run_ID links their companion tables. The ingestion result establishes compatibility for these recorded outputs. Prospective extensibility is a separate question. A follow-up study could add previously unseen run types and check whether the existing ingestion and analysis code accepts them unchanged. Outputs with different meanings or structures may require changes to an adapter or the schema.

For RQ2, the performance and timing plots show how the retained dataset identifiers allow comparisons within a common task and release. The individual and cumulative resource criteria identify which recorded experiments meet a stated constraint, either across the full collection or within a selected version. Such a criterion describes the experiment under its recorded execution conditions. A 15-minute training limit, for example, concerns the cost of fitting a model. It gives no direct assessment of the latency or false-alarm burden of an operational detector. Those deployment questions require measurements and evaluation conditions appropriate to the intended network, as discussed by Sommer and Paxson [11].

Energy estimates require a further qualification. Train_Energy_Wh combines the assigned device TDP with observed CPU utilisation and training duration. It is a model-based approximation, without calibration against RAPL readings or an external power meter in this experiment. Bouza et al. describe how energy-accounting tools differ in their treatment of hardware and utilisation [20], while Green Algorithms makes the assumptions required for estimating computation-related emissions explicit [19]. Here, the 5 Wh criterion expresses a threshold on the recorded proxy. Its relationship to measured electricity consumption remains untested. Inference-energy coverage also varies across implementation paths, which is why the reported energy screening uses training estimates.

Predictive comparisons are limited to a single stratified partition and the configurations executed on the CPU profile. Imputation uses a source-sample map prepared before the final split, whose construction does not enforce training-only membership. Rows later assigned to the test set can therefore influence the imputation values. The recorded outputs do not quantify the effect on predictive scores. A follow-up evaluation should fit imputation after assigning the training partition. Differences between models also include their configured architectures and training procedures, so the Small, Medium, and Large labels alone cannot isolate an effect of the software library. The collection contains results from several datasets without testing transfer between them. Repeated partitions and explicit transfer experiments would be needed to examine the stability of model comparisons and their applicability across network environments.

A larger evaluation could test whether the model configurations selected under a resource budget remain preferable across repeated splits and hardware profiles. Direct power measurement would allow the error of the TDP approximation to be quantified, following the measurement-oriented approach exemplified by MLPerf Power [16]. The existing run identifiers would then need to distinguish each repetition and execution environment, while retaining the configuration information needed to compare them. Such experiments would extend the empirical conclusions beyond the current demonstration of ingestion and descriptive analysis.

## 7 Conclusion

For the 216 recorded runs, GreenNIDS represents heterogeneous NetFlow model outputs in one analysis schema and links summary measurements to their companion tables. The same collection supports comparisons of MCC and training time within dataset–task groups, followed by screening under specified resource limits. Across all three versions, 25.0% of runs satisfy the joint MCC, training-time, peak-RAM, and estimated training-energy criteria. These are results for the recorded preprocessing and CPU profile, with energy expressed through a TDP-based estimate. The demonstrated contribution is the ability to change groupings and screening criteria after execution while retaining the measurements and configurations behind each comparison.